# Testing the Least Action Principle in an $\Omega_\circ = 1$ Universe

E.Branchini[1] and R.G.Carlberg[2]

[1] SISSA - International School of Advanced Studies,

Via Beirut 2, I-34014 Trieste, Italy.

e-mail address: enzo@tsmi19.sissa.it

[2] Department of Astronomy, University of Toronto,

Toronto Ontario Canada M5S 1A1.

# ABSTRACT

The least action principle (LAP) is a dynamically rigorous method for deriving the history of galaxy orbits. In particular it is an $\Omega$ test, predicting current epoch galaxy velocities as a function of position and of the cosmological background. It is most usefully applied to in-falling structures, such as the local group, where its application indicates that the preferred cosmological model is $\Omega_\circ = 0.1$ and $h = 0.75$ ($h$ is the Hubble parameter in units of 100 Km s$^{-1}$ Mpc$^{-1}$). The method assumes that all the mass acts as if it were distributed as the visible galaxies. We test the reliability of the LAP to Local Group-like systems extracted from $\Omega_\circ = 1$ $n$–body simulations. While the orbits of the galaxies are qualitatively well reconstructed, the LAP systematically underestimates the mass of the system. This failure is attributed to the presence of extended halos weakly clustered around visible galaxies which prevent a large fraction of the group mass from being detected by the LAP technique. We conclude that the LAP method cannot rule out an $\Omega_\circ = 1$ value on the Local Group scale. Better constraints on $\Omega_\circ$ may be obtained by applying this technique to in-falling systems, such as clusters, containing objects with separations large compared to galaxy sizes.

*Subject headings*: Cosmology – Galaxies: clustering – Galaxies: interactions

1. INTRODUCTION

Dynamical estimates of the mean density of the Universe display an apparent contradiction. On scales $< 50\ h^{-1}$ Kpc the Milky Way satellites' dynamics (Faber & Gallagher 1979; Little & Tremaine 1987; Zaritsky *et al.* 1989), the timing argument (Khan & Woltjer 1959; Gunn 1975; Gott & Thuan 1978; Mishra 1985; Einasto & Lynden-Bell 1982 and Raychaudury & Lynden-Bell 1989) and the internal dynamics of spiral galaxies (Salucci *et al.* 1993) are incompatible with an $\Omega_o = 1$ value. Similar indications come from larger scale density determinations based on the application of Cosmic Virial Theorem (Peebles 1976) on the pairwise velocity dispersion of galaxies (Bean *et al.* 1983; Davis & Peebles 1983; Hale-Sutton *et al.* 1989) and the virial estimation of galaxy cluster masses. The bulk of the evidence indicates a low $\Omega_o$-value ($\leq 0.35$) on scales $< 10\ h^{-1}$ Mpc. On the other hand, dynamical tests carried out on scales larger than $\sim 30\ h^{-1}$ Mpc give different results. The IRAS galaxies' dipole moment (Rowan-Robinson *et al.* 1990; Strauss *et al.* 1992), and the comparison of the IRAS galaxies' radial peculiar velocity field with the predicted velocities obtained from the observed galaxy number density (Strauss & Davis 1988; Kaiser *et al.* 1991) and the IRAS/POTENT method (Dekel *et al.* 1993), are both consistent with an $\Omega_o = 1$ Universe, possibly indicating that $\sim 90\%$ of the mass in the Universe clusters on scales larger than $10\ h^{-1}$ Mpc (Peebles 1993).

Most of the $\Omega_o$ measurements on small scales use virialized systems where the galaxy velocity field is assumed to trace from the velocity field of the mass distribution. Several $n$–body experiments find that the pairwise random velocity of galaxies on small scales is a factor 2 lower than that of the dark matter (Carlberg & Couchman 1989; Carlberg, Couchman & Thomas 1990; Carlberg 1991; Couchman & Carlberg 1992; Gelb & Bertschinger 1993), implying that the $\Omega_o$–value deduced through the Cosmic Virial Theorem should be corrected upward by a factor of 4. In addiction the single galaxy velocity bias has been constrained in the range $0.7 - 0.9$ (Carlberg & Dubinski 1991; Katz & White 1993; Carlberg 1993) and the cluster virial masses indicated by the galaxies are $15 - 20\%$ of the true mass. Moreover, if galaxies are biased tracers of the mass density field, then hydrodynamical effects may increase the value of the density bias on small scales (Cen & Ostriker 1992).

The Local Group (hereafter LG) offers a unique opportunity to constrain $\Omega_o$, being generally regarded as a dynamically young system in which galaxies, whose distances are reasonably well known, are in–falling for the first time. Under such conditions the velocity bias is expected to be negligible (Carlberg 1991). The LG is essentially a binary system composed by two nearly equally massive galaxies (the Milky Way and M31) surrounded by dwarf companions (N6882, I1613, WLM, Sex A, N1309 and N300). Different techniques for estimating the mass inside the LG lead to a range of results. Techniques based on the satellites' dynamics constrain the mass contained in our galaxy within a radius of $50 - 100 \ h^{-1}$Kpc to the range $2 - 10 \times 10^{11} \ M_\odot$.

On the other hand, the LG mass derived using the timing argument is a factor 2–3 larger. The first value obtained by Khan and Woltjer (1959) ($\simeq 1.8 \times 10^{12}$ $M_\odot$) has been enhanced by a factor $\sim 3$ in subsequent analyses. As pointed out by Kroeker and Carlberg (1991), since the timing argument mass determination depends on the system size then the LG mass is consistent with an $\Omega_o = 1$ Universe. Peebles (1989, hereafter P1) has proposed a method based on the least action principle to reconstruct the trajectories of the LG galaxies back in time. With this technique it is possible to study the effect of the external masses on dynamics of the Local Group and to determine its mass. The influence of the external tidal field can be larger than the Raychaudury and Lynden-Bell (1989) estimate: the LG mass increases to $6.4 \times 10^{12}$ $M_\odot$ by considering the mass of the two nearest groups, and to $\sim 1.9 \times 10^{13}$ $M_\odot$ taking into account all the groups within a sphere of $\sim 6$ $h^{-1}$ Mpc centered on the Milky Way (Dunn & Laflamme 1993). Furthermore, the LAP method can constrain the amplitude of the primordial mass fluctuations responsible for the observed local galaxy distribution (Peebles 1990, hereafter P2). Finally, the LAP can be used to discriminate among different cosmological models by analyzing the dynamics of the two neighboring Maffei and Sculptor groups (P1 and P2). According to the McCall's (1989) distance determination to I342, the Maffei group is receding with a large positive line–of–sight peculiar velocity that no cosmological model, coupled with the LAP technique, is able to reproduce. Similar problems on smaller scales are related to the N6822 measured distance. If one accepts the P1 assumption that Maffei and Sculptor group distances are comparable ($= 2.66$ $h^{-1}$ Mpc), then those two groups have a small negative peculiar velocity with respect to the Milky Way. The LAP method shows that a flat cosmology with $\Omega_\circ = 0.1$ and $h = 0.75$, correctly reproduces approaching trajectories, while in an $\Omega_\circ = 1$ Universe Maffei and Sculptor groups, being in a local density void, would recede from the LG.

The purpose of this paper is to test the reliability of the LAP method in an Einstein–De Sitter Universe. In the following section, the LAP is implemented using the hypotheses that the mass in the Universe is now concentrated around visible galaxies and that, at early epochs, protogalaxy masses were in regions with simple boundaries. Under these assumptions it is possible to concentrate the mass within each region such that the gravitational field at the neighboring positions approximates that of the true mass distribution. This assumption is a concern in an $\Omega_\circ = 1$ Universe where galaxies at the present epoch are likely to be surrounded by extended *overlapping* halos. To test this hypothesis we compare the mass and the orbits determined using this technique with the analogous quantities measured in $\Omega_\circ = 1$ CDM $n$–body simulations. In Section 3 we analyze one galaxy cluster simulation in which the number of particles per galaxy is high, making it possible to compare directly the LAP orbits to the galaxy trajectories observed in the $n$–body experiments. In Section 4 we extract a mock catalogue of LG candidates from a cosmological simulation and the LAP method is used to measure the masses of the selected groups. The last section contains a discussion of the results.

## 2. THE LEAST ACTION METHOD

Peebles' implementation of the least action principle assumes that the peculiar gravitational acceleration of a galaxy can be estimated by concentrating the LG mass around the visible galaxies. The LG can therefore be regarded as a system of point-like objects labeled by $i$, with mass $m_i$ at the comoving position $\mathbf{x}_i(t)$ at the time $t$, and with softening parameter $\epsilon_i$ associated with the physical extension of the galaxies (P1).

In the hierarchical clustering picture small-amplitude fluctuations in a nearly homogeneous Universe that grow to form protogalaxies have to be such that the central positions at the present epoch, $\mathbf{x}_i(t_o)$, coincide with the observed galaxy positions. The classical orbits that correspond to minima for the action $S$ can be completely specified by fixing the three components of the observed positions and by requiring the three peculiar velocity components to vanish at early times:

$$\delta \mathbf{x}_i = 0 \quad \text{at} \quad t = t_o, \quad \text{and} \quad a\frac{d\mathbf{x}_i}{dt} \to 0 \quad \text{when} \quad a \to 0, \quad (2.1)$$

where $a$ is the scale factor and $a_o = a(t_o) = 1$.

The equations of motion can be derived from the action

$$S = \int_0^{t_o} \left[ \sum \frac{m_i a^2}{2} \left(\frac{d\mathbf{x}_i}{dt}\right)^2 + \frac{G}{a} \sum_{i \neq j} \frac{m_i m_j}{|\mathbf{x}_i - \mathbf{x}_j|} + \frac{2}{3}\pi G \rho_b a^2 \sum m_i \mathbf{x}_i^2 \right] dt, \quad (2.2)$$

where $\rho_b$ is the mean background density.

The present work is aimed at testing the LAP technique in an Einstein–De Sitter Universe. Nevertheless, a low density cosmology will also be considered. In this case we will use a non–vanishing cosmological constant both to satisfy the flatness requirement of the inflation scenario and to bring the age of the Universe in range of the currently accepted values. The time interval is thus

$$H_\circ dt = \frac{a^{1/2} da}{[\Omega_\circ + a^3(1 - \Omega_\circ)]^{1/2}}. \tag{2.3}$$

To find the numerical approximation to the orbits which minimizes the action, we use the following parametric trial functions:

$$x_i^\alpha(t) = x_i^\alpha(t_o) + \sum_n C_{i,n}^\alpha f_n(t), \tag{2.4}$$

where $\alpha = 1, 2, 3$ denote each spatial component of the orbits, the $f_n(t)$ are functions satisfying (2.1) and $x_i^\alpha(t_o)$ are the final positions of the galaxies. In this work we take $f_n = a^n(1-a)$ for $n = 0, 1, 2, 3, 4$. As already pointed out in P2, using other simple trial functions do not affect appreciably the final result.

Using (2.2) and (2.4), the derivative of the action with respect to the unknown coefficients $C_{i,n}^\alpha$ is

$$\frac{\partial S}{\partial C_{i,n}^\alpha} = m_i \int_o^{t_o} f_n(t) \left[ \frac{4}{3}\pi G \rho_b a^2 x_i^\alpha(t) - \frac{d}{dt} a^2 \frac{dx_i^\alpha}{dt} + \frac{G}{a} \sum_j \frac{x_j^\alpha - x_i^\alpha}{|\mathbf{x}_i - \mathbf{x}_j|^3} \right] dt. \tag{2.5}$$

The orbits corresponding to the constraints

$$\frac{\partial S}{\partial C_{i,n}^{\alpha}} = 0,$$

satisfy the equations of motion which, using (2.3), can be written as:

$$a^{1/2}\frac{d}{da}a^{3/2}\frac{d\mathbf{x}_i}{da} + \frac{3(1-\Omega_\circ)a^4}{2F}\frac{d\mathbf{x}_i}{da} = \frac{\Omega_\circ}{2F}\left[\mathbf{x}_i + \frac{R_o^3}{M_T}\sum_j \frac{m_j(\mathbf{x}_j - \mathbf{x}_i)}{|\mathbf{x}_j - \mathbf{x}_i|^3}\right],$$
(2.6)

where $F = \Omega_\circ + (1-\Omega_\circ)a^3$ and $R_o$ is the radius of the sphere that would contain the LG total mass $M_T$ in a homogeneous Universe at the present time.

To minimize the action we follow the gradient with increments

$$m_i \delta C_{i,n}^{\alpha} \propto -\frac{\partial S}{\partial C_{i,n}^{\alpha}},$$
(2.7)

starting from $C_{i,n}^{\alpha} = 0$ and using 60 uniformly spaced steps in $a$ for the numerical evaluation of $\partial S/\partial C_{i,n}^{\alpha}$. We also replace the inverse square law by using the same softening parameter $\epsilon_i = 0.03 \times R_o$ for every galaxy. The method was tested by reproducing the results in P1. We note that this technique cannot be applied when the crossing time $t_c$ is much less than the Hubble time $t_0$, as in this case the LAP solution is not unique. A dynamically young system is therefore required for the method to find the orbits without ambiguity.

We tested the LAP method on the data extracted from $n$–body simulations. The galaxies are identified using the procedures described in next sections. Each galaxy is characterized by its center of mass position and velocity and by its total mass. The LAP method is then applied by assuming fixed the galaxy mass ratios and leaving the total mass of the system as a free parameter. The value of the system mass, which we will refer to as $M_{LAP}$, is finally adjusted to match the radial velocity between the two largest bodies, as measured in the $n$–body simulation.

## 3. HIGH RESOLUTION N–BODY TEST

The data were obtained from the recent Carlberg (1993) collisionless $n$–body simulation designed to mimic the evolution of a galaxy cluster with enough particles to resolve galaxy scales. A sphere of 12.5 $h^{-1}$ Mpc comoving radius containing a peak extracted from a CDM $\Omega_\circ = 1$ Universe realization was initialized with $N = 10^6$, particles each having a mass $2.27 \times 10^9$ $h^{-1}$ $M_\odot$. The gravitational evolution of the system was followed using a multistep quadrupole TREECODE (Barnes & Hut 1986, Dubinski 1988), taking into account the effect of an external quadrupole tidal field. The simulation employed a softening length $\epsilon_1 = 7.8 h^{-1}$Kpc and a minimum timestep of 1/1264 of a Hubble time. The inverse of the (linearly extrapolated) mass variance at $r = 8$ $h^{-1}$ Mpc, $b_8$, was unity at $z = 0$. The problem of galaxy identification in this simulation is greatly simplified by the existence of dense pieces of long–lasting self–gravitating substructures with a velocity dispersion of $\sim$ 100 Km s$^{-1}$. This feature guarantees the stability of our results when different galaxy identification procedures are used.

At the final epoch the cluster is a virialized triaxial anisotropic object with half mass radius $r_{1/2} = 1.13\ h^{-1}$ Mpc and a one dimensional RMS velocity dispersion of 1080 km s$^{-1}$. As a whole, this system is not suitable for testing the LAP method reliability because in its central part $t_c < t_0$. To obtain a dynamically young LG–like system, the original $n$–body data were transformed as follows: we considered the same cluster at an earlier epoch keeping constant the total mass of the system and then the dimensions of the system were scaled to match the LG size. This transformation preserves both the global density and the form of the equation of motion. In practice we considered the simulated cluster at t=17.8 and we scaled it to obtain a group in which galaxies are falling for the first time. The new system is still composed of $10^6$ particles but now has a comoving radius of 2.5 $h^{-1}$ Mpc. The total mass contained in the simulation, $M_T$, turns out to be $1.8 \times 10^{13}\ h^{-1}\ M_\odot$, which is used as the total mass of the group. This transformation results in two problems which cause the final system to differ from a typical group of galaxies in an $\Omega_o = 1$ Universe. Firstly, the original quadrupole tidal field was computed on a 12.5 $h^{-1}$ Mpc comoving radius sphere, while the linear dimension of the new system is a factor of 5 smaller. This causes the tidal field to be slightly underestimated. The second, more serious, incongruity refers to the galaxies' halos: scaling the system to the LG dimensions reduced the halo–extension, which is a fundamental feature of an Einstein–De Sitter Universe. As a result the rescaling can affect the results of the LAP-method reliability tests by reducing the dynamical influence of the halos. We will quantitatively estimate this effect in Section 4 by carrying out a different analysis.

To identify the galaxies, we used two procedures. In the first one we selected all the groups at a percolation length $\alpha \overline{d}(t)$, where $\overline{d}(t)$ is the mean inter-particle separation at time $t$. The overdensity associated with the corresponding peaks is $\frac{\delta \rho}{\rho} \propto (\alpha \overline{d}(t))^{-3}$. All the peaks above this threshold containing a specified minimum number of points have been recognized as galaxies. We applied this procedure using several different overdensity thresholds in order to test the dependence of the LAP solution on the $\frac{\delta \rho}{\rho}$ used. A second two–step procedure was also applied. In this case we identified the protogalaxies at $t = 6$ with groups containing at least 10 particles, found at a percolation length of $0.2 \times \overline{d}(t)$. These particles were relinked at $t = 17.8$ using a percolation length of $0.05 \times \overline{d}(t)$, corresponding to a minimum $\frac{\delta \rho}{\rho} = 8000$ (see Carlberg & Dubinski (1991) for more details on the link algorithm). The aim here was to test the possible dependence of the results on the adopted galaxy identification procedure. Finally the LAP technique was applied to reconstruct the orbits of the largest selected galaxies.

In Table 1 we report, for each experiment, the overdensity threshold $\frac{\delta \rho}{\rho}$, the number of peaks considered $N$, and the ratio $M_T/M_{LAP}$. To be consistent with the $n$–body background cosmology we used $\Omega_o = 1$ and $h = 0.5$.

Figure 1 shows the time evolution of a typical peak identified at a $\frac{\delta \rho}{\rho} = 8000$ overdensity threshold using the two–step procedure. The high merging rate causes the motion of the center of mass to poorly describe the history of a galaxy orbit within the group. Peaks identified using different criteria evolve in a similar way.

In figure 2a and 2b the LAP orbits of galaxies with $\frac{\delta\rho}{\rho} = 8000$ and $\frac{\delta\rho}{\rho} = 128$ respectively, are compared with their true center of mass trajectories obtained from $n$–body data. In both cases a general qualitative agreement is found. The LAP orbits for bodies 6 and 7 in figure 2a, as well as for body 12 in figure 2b, however, are significantly different from the true trajectories. Moreover, the orbits of bodies 3 and 4 in figure 2a are exchanged (although in this case we recovered the correct solution by slightly varying the density threshold). It is unclear if in these cases the center of mass motions give misleading indications of the peaks' dynamics or, more probably, if the correct orbits represent stationary points which are not minima for the action (analogously to the N6822 case in P2).

The second main result, shown in Table 1, is that the $M_{LAP}$ is systematically lower than $M_T$. The average $M_T/M_{LAP}$ value is 1.6 with $\sigma = 0.27$. The stability of the mass ratio in all the different experiments indicates that this result depends neither on the galaxy identification procedure, nor on the overdensity threshold chosen. The magnitude of the mass discrepancy depends on the definition of the system's mass. In this case the quoted $M_T/M_{LAP}$ value should be regarded as a lower limit since, as we have already pointed out, the adopted rescaling hides possible failing which would otherwise occur int this case.

Figures 3a and 3b compare the true peak velocity $v_{TRUE}$ from the $n$–body data with the same quantity found as a LAP solution ($v_{LAP}$). The same experiments as in figures 2a and 2b are considered. The underestimation of the mass causes $v_{LAP}$ to be larger than $v_{TRUE}$. Interestingly the mass needed to match the peculiar velocity of the outer galaxies turned out to be very similar to $M_T$.

In addition to the previous analyses, we have also performed the following tests:

*a) The effect of varying the number of orbits modelled $N$.*

After considering the peaks selected at an overdensity threshold $\frac{\delta\rho}{\rho} = 8000$ with the two–step procedure, we applied the LAP technique using four different values of $N$: 8, 10, 14 and 17. The solution was not affected by changing this parameter: the same qualitative agreement between the LAP and $n$–body orbits was found, and no significant variations in the ratio $M_T/M_{LAP}$ were observed ($= 1.75 \pm 0.02$). This $N$–independence arises because most of the system mass resides in the three largest bodies and thus variations in the number of light bodies considered do not affect appreciably the system's dynamics.

*b) The effect of varying the final epoch.*

Since the two largest galaxies in the simulation are surrounded by extended overlapping halos, one might suspect that their relative velocity is biased with respect to the dark component, even if they are approaching for the first time. The mass determination procedure would thus underestimate the total mass of the group. To test this hypothesis we considered the system at an earlier epoch, when the bias velocity is expected to be negligible, or at least smaller than at $t = 17.8$. An overdensity threshold $\frac{\delta\rho}{\rho} = 8000$ was considered to identify galaxies at $t = 10.5$. We found that $M_T/M_{LAP} = 1.45$, only a factor 1.1 smaller than the average of values reported in Table 1. Thus, the velocity bias in the relative motion of the two largest bodies at $t = 17.8$, if present, has to be very small and cannot be the cause of the observed mass underestimate.

*c) The effect of varying the cosmological parameters $h$ and $\Omega_o$.*

P2 has shown that the internal LG dynamics is nearly independent of $\Omega_o$. As the density parameter, however, strongly influences the motion of the neighboring groups, the LAP method was used to constrain the value of $\Omega_o$. In this dynamical $\Omega_o$ test a hidden $h$–dependence is present because the distance of external galaxies is based on the infrared Tully–Fisher relation which is calibrated to the redshift of more distant galaxies (Peebles 1988). In practice this causes the LAP $\Omega_o$ determination to be model dependent. As an example, P2 chose $(\Omega_o, h)$ by constraining the expansion time scale $t_o$ in the range of currently accepted values and he allowed for a non–vanishing cosmological constant to obtain $t_o = 17$ Gyr in his $\Omega_o = 0.1$, $h = 0.75$ preferred model.

In our case the cosmological background is known since we are considering $n$–body data. Nevertheless, as shown in figures 3a and 3b, the LAP method fails to reproduce the correct peculiar velocities. An interesting test is to apply the LAP method varying $(\Omega_o, h)$ in order to explore if an incorrect cosmological model can bring the peculiar veloc-ities closer to the correct values. We considered the 10 largest galaxies selected at a $\frac{\delta\rho}{\rho} = 8000$ threshold and we added six peaks containing more than 100 particles and having a distance from the center of mass of the group in the range $4.2 - 4.8$ $h^{-1}$ Mpc. To reproduce the observational uncertainties we introduced an $h$ dependence in the outer peaks' distances. Using the correct cosmological parameters ($h = 0.5$ and $\Omega_\circ = 1$) we found that the average $v_{LAP}/v_{TRUE}$ value for these peaks is 1.45. We reanalyzed the same n-body data, but considering it to be an $\Omega_\circ = 0.1$, $h = 0.75$ flat Universe (with $\Lambda = 0.9$). The presence of a cosmological constant weakly influences the dynamics within the group but there is no appreciable difference between the predictions of an open model and a cosmological flat one with the same values of $\Omega_o$ and $t_0$ (Peebles 1994). Analogously, the dynamics of the central part of the group was not appreciably influenced by introducing small external objects, or by changing the $\Omega_\circ$-value. However, in the low density flat model, the value $v_{LAP}/v_{TRUE}$ averaged over the outer galaxies turned out to be 0.83. Thus, paradoxically, the LAP method gives similar or even better results when an incorrect cosmological model is used. This indicates that, if $\Omega_\circ = 1$, the LAP technique cannot reliably determine the density parameter value in the LG neighborhood.

*d) The effect of varying the softening parameter $\epsilon_i$.*

Extended galactic halos can be partially modelled by altering the $1/r$ potential to $1/\sqrt{r^2 + \epsilon_i^2}$, where $\epsilon_i$ is a softening parameter. After taking galaxies selected at a $\frac{\delta\rho}{\rho} = 8000$ threshold, we applied the LAP technique using several different $\epsilon_i$–values. The original $\epsilon_i = 0.03 \times R_o$, corresponding to a value of 200 Kpc, was chosen in previous works (P1 and P2) to take into account the mass distributed in the halos. We increased the softening by a factor 2 and found that neither the orbits nor the $M_T/M_{LAP}$ ratio were greatly changed. Very similar results were obtained when $\epsilon_i$ was set equal to 0, i.e., point–like galaxies. As already pointed out by P2, the observed $\epsilon$–independence for $\epsilon_i$ in the range $0 - 400$ Kpc, derives from the fact that the softening parameters has a local effect, while the galaxy orbits found in the LAP solution are seldom close.

In an $\Omega_o = 1$ Universe galaxy halos can be larger ($1 - 2$ $h^{-1}$ Mpc). It is interesting to explore if the LAP method can take such extended halos into account. Since in the LAP implementation the galaxy extension is quantified by the softening parameter, the question can be modified as follows: how large should the softening parameter be in order to reproduce the dynamical effect of extended halos ? To answer this question we reran the same LAP simulation with several different softening parameters until we obtained $M_T/M_{LAP} = 1$. The $n$–body experiment indicated that galaxy halos have a constant extension in physical coordinates back to $z = 1.15$. Since $\epsilon_i$ is expressed in comoving coordinates, we modeled a time–dependent softening parameter $\epsilon_i(t)$ to keep its value constant in physical coordinates back to $z = 1.15$. At earlier times the softening was fixed equal to $\epsilon_i(z = 1.15)$. The present–time $\epsilon_i$–value for which $M_T = M_{LAP}$ was $\sim 1$ Mpc. This large time–dependent softening parameter causes the external bodies to be more bound, resulting in a better match to their $n$-body peculiar velocities. However, the LAP-orbits of all the galaxies become shorter than the true ones. In spite of the larger, time dependent softening parameter, the LAP solution is still different from the $n$-body. These results can be accounted for if, as we will verify in the next section, galaxies are embedded in extended haloes which act as a hot mass component on the scale of the galaxy group. In this scenario the peculiar gravitational acceleration of galaxies whithin the group is determined by the mass concentrated around visible objects. A small softening parameter is therefore sufficient to reproduce the motion of inner galaxies. However, a large fraction of the mass is distributed in a weakly clustered intragroup background which causes the outer galaxies to have systematically smaller peculiar velocities, as observed. Therefore it is not surprising that a large softening parameter cannot both alleviate the missing mass problem and reproduce the motion of galaxies within the group.

### 4. LOW RESOLUTION N–BODY TEST

In this test the data were obtained from Couchman and Carlberg's (1992) cosmological $n$-body simulations. In this case the background cosmology is also an $\Omega_\circ = 1$, $h = 0.5$, CDM model. Two computational cubes with sizes of 100 and 200 $h^{-1}$ Mpc respectively, were considered. The total number of particles in each simulation is $128^3$. Here we analyze the 100 $h^{-1}$ Mpc cube data, thus the particle's mass is $2.65 \times 10^{11}$ $M_\odot$ ($h = 0.5$). $b_8 = 1$ after 305 equal timesteps. The equivalent softening parameter for a Plummer potential corresponds to 15 $h^{-1}$ Kpc, constant in physical units. Since no rescaling has been applied to the data, we avoid tidal field and halo–extension problems discussed in Section 3.

We adopted the galaxy identification procedures described in the previous section. In the first analysis we selected all the peaks at the final epoch, using the previous link method, at three different overdensity thresholds ($\frac{\delta\rho}{\rho} = 125$, 8000 and 64000) containing at least 5 particles. The second two–step procedure identified, after 50 timesteps, all the peaks having $\frac{\delta\rho}{\rho} \geq 125$, and then relinked the selected objects at the final epoch, using the same criteria followed in the former selection.

The next step was to consider binary galaxies resembling the Milky Way–M31 system to construct fictitious catalogues. We started by considering only objects having masses in the range $1.32 - 3.97 \times 10^{12}$ $M_\odot$ (i.e. peaks containing $5 - 15$ points). This is approximately the mass of our Galaxy when determined within a radius of $\sim 100$ Kpc, which turned out to be the typical radius of the sphere containing a $5 - 15$ points peak in the $\frac{\delta\rho}{\rho} = 8000$ threshold experiment.

The algorithm for finding binaries was: for each galaxy;

1. Find the closest neighbor and its distance $r_1$.

2. Find the next closest galaxy to the center of mass of the binary system and its distance $r_2$.

3. Compute the ratio $r_2/r_1$.

At this point the following constraints were applied:

a. $0.3 < r_1 < 2$ (in Mpc units). The lower limit was imposed to avoid interacting galaxies in the sample. The upper limit was required to obtain a system similar to LG.

b. $r_2 \geq 5$ Mpc, since the observed nearest groups are at a distance of 5.32 Mpc ($h = 0.5$) from the LG.

c. Only approaching binaries were considered.

d. $r_2/r_1 \geq 2.5$, in order to eliminate binaries which are members of a cluster.

One more condition, based on the total energy of the system, was applied. The sign of the energy can be calculated using the formula for a two–body system

$$\frac{E}{M_1 M_2} = \frac{v^2}{2M} - \frac{G}{r_1}, \qquad (4.1)$$

where $E$ is the total energy of the system, $M_1$ and $M_2$ are the masses of the individual galaxies, M is the total mass of the system, whose definition is given below, and $v$ is the total relative velocity. Our purpose was to select LG candidates, but the question of whether the Milky Way–M31 system is bound or not does not have a clear answer because of the lack of information on the relative transverse velocity $v_t$. For example, according to P2 and the Dunn & Laflamme (1993) results, the Milky Way and M31 fail to be bound because of the influence of the external groups. However, if $v_t$ were 0 then the system would be bound. Therefore, we replaced $v$ with the radial velocity $v_r$ in (4.1), and we required the l.h.s. to be negative.

As in the previous analysis, the magnitude of the possible discrepancy between $M_{LAP}$ and the true mass depends on the definition adopted for the total mass of the system. We know from the work of Kroeker & Carlberg (1991) that the timing mass for approaching isolated binaries is well approximated (in an $\Omega_\circ = 1$ CDM $n$-body simulation) by the mass contained into two spheres centered on the galaxy positions and with radius equal to $r_1/2$ ($M_K$). Moreover, for first–approach binary systems, the LAP method gives a similar result to the timing argument ($M_K \sim M_{LAP}$). Here we use two different definitions for the total mass of the system:

i) $M_5$ is the mass contained within a sphere of 5 Mpc radius centered on the center of mass of the binary system, as measured from the $n$-body data. This is analogous to the $M_T$ definition previously adopted.

ii) $M_r$ is the mass contained within a sphere of radius $r = r_2 - 0.5$ Mpc, centered on the center of mass of the system. The 0.5 Mpc gap between $r$ and $r_2$ has been introduced to avoid possible contamination from the outer objects' halos.

If the mass of the system were actually concentrated around the seen galaxies, then $M_K$, $M_{LAP}$, $M_5$ and $M_r$ would coincide.

For each binary we computed the crossing time and the free fall crossing time $t_{ffc}$,

$$t_c = \frac{r_1}{|v_r|}, \qquad t_{ffc} = \frac{\pi}{2}\sqrt{\frac{r^3}{2GM}}, \qquad (4.2)$$

where $M$ can be either $M_5$ or $M_r$. Then, we divided the selected binaries into three different categories. The first one contains "wide" binaries, defined as those systems having both $t_c$ and $t_{ffc}$ larger than $t_o$. In the second category are the "quasi–wide" systems having

$$\frac{t_o}{2} < t_c < t_o \qquad \text{and} \qquad \frac{t_o}{4} < t_{ffc} < \frac{t_o}{2}.$$

All other binaries have been included in the third category. Since these last systems are likely to have already completed at least one orbit, the relative LAP solution is not unique. Therefore they will be included only for reference. Finally we applied the LAP method to find the masses of the selected binaries.

Results of LAP experiments performed using the different binary catalogues are summarized in Table 2. The first column contains the experiment reference number and the second the overdensity threshold. In the remaining columns each row is divided in three subrows to take into account the three different categories defined above. The first subrow contains the parameters of "wide" binaries only. In the second, data from the first two categories are listed, while in the last all the objects are included. Columns 3, 4 and 5 contain the number of binaries found, the average value of $\log(M_5/M_{LAP})$ and the relative standard deviation respectively for the $M_5$ case. Columns 6 to 8 contain the same parameters for the $M_r$ definition. The last three columns contain the number of best LG candidates and the corresponding average values of $\log(M_5/M_{LAP})$ and $\log(M_r/M_{LAP})$. The best LG candidates are constrained to have $r_1$ and $|v_r|$ in the range $600 - 850$ Kpc and $100 - 150$ Km s$^{-1}$ respectively.

In this analysis the results depended slightly on both the galaxy identification procedure and the overdensity threshold. We mainly considered results from the catalogues containing galaxies selected at $\frac{\delta\rho}{\rho} = 8000$ since it was found that peaks above this overdensity threshold are good galaxy tracers (Carlberg & Dubinski 1991; Couchman & Carlberg 1992). In all the performed experiments the LAP method underestimated the system's mass. The average $M_5/M_{LAP}$ ratio is $\sim 5$, significantly larger than in the previous analysis. As in the last section this mass underestimate suggests an intrinsic failure of the LAP method, presumabably related to the assumption that mass is concentrated around galaxies. To test whether mass is strongly clustered around density peaks or distributed in extended halos, we measured the mass contained in spheres of increasing radius centered on the midpoint of the binary system, normalized to the mass contained in a sphere of radius $r_{min} = r_1/2 + 0.2$ Mpc. The same procedure was repeated for each binary in the catalogue considered. Though the details depend on the selection criteria, we have found a result which is common to all catalogues considered: the mass is preferentially located on extended halos weakly clustered around the binary system. This is clearly shown in figures 4a and 4b, in which we have plotted the quantity $M(<R) = \frac{\int_o^R (\rho(r)-\rho_b) d^3r}{\int_o^{r_{min}} (\rho(r)-\rho_b) d^3r}$ averaged over all the binaries of the two different catalogues. This is the mass distribution one would expect to find in a biased galaxy formation scenario.

## 5. DISCUSSION AND CONCLUSIONS

The two different analyses carried out in Sections 3 and 4 indicate a substantial failure of the LAP method to describe the dynamics on LG scales in an $\Omega_\circ = 1$ Universe. In Section 3 we applied the LAP method to a group of galaxies obtained by transforming the data from an $n$–body simulation. Inconsistencies related to the $n$–body data rescaling reduce the extension of the halos associated with the member galaxies, therefore this test will not reveal the full extent of the problem with using the LAP technique in an $\Omega_\circ = 1$ Universe. The LAP method succeeds in qualitatively reconstructing the galaxy trajectories. This result is not obvious since the high merging rate and the presence of large halos surrounding the objects cause the center of mass displacement to be a poor tracer of the group's dynamics. This success, however, appears less striking under a more quantitative investigation. The mass of the system determined using the LAP method turns out to be systematically underestimated. The quoted $M_T/M_{LAP}$ ratio is $1.6 \pm 0.23$. Although the magnitude of the discrepancy depends on the definition adopted for the mass of the system, the effect is undoubtedly real as indicated by the motion of external members of the group. This result does not depend on the peak identification procedure. Table 1 shows that varying the overdensity threshold by a factor $10^4$ does not seriously affect the $M_T/M_{LAP}$ ratio. This behavior was in part expected, since the peaks in the analyzed simulation constitute a well–defined, long–lasting population of objects, easily identified by linking algorithms using different linking lengths. Tests performed to find the origin of this discrepancy indicate that the LAP solution depends neither on the number of peaks considered nor on the softening parameter chosen. Similarly we have found that the velocity bias in the simulation is very small and does not appreciably influence the $M_{LAP}$ determination.

In Section 4 we applied the LAP method to fictitious catalogues of binary galaxies extracted from a large volume $n$–body experiment. No rescaling was applied to the data. After adopting two distinct system mass definitions, we found that the mass was still underestimated and that the mass discrepancy was larger than in the previous test, as predicted in Section 3.

The bulk of the evidence indicates that the origin of the $\Omega_0$ underestimate has to be found in the LAP method itself. In Sections 1 and 2 we underlined that one of the key assumptions in the LAP method's implementation is that all the matter in the Universe is concentrated around visible galaxies. This approximation is well justified in an $\Omega_\circ = 0.1$ Universe, in which the galaxies are surrounded by halos extending up to 200 Kpc. However, as we have verified, our $\Omega_\circ = 1$ CDM $n$–body simulation is characterized by the presence of much more extended overlapping halos weakly clustered around the peaks of the density field, as expected in a biased galaxy formation picture. Since the LAP technique can only reveal the mass clustered around visible objects, a dependence of the mass discrepancy on the relative separation of galaxies would be expected. In figure 5a and 5b we have plotted the mass discrepancy $\log(M_5/M_{LAP})$ versus the galaxy separation $r_1$ for binaries of two differ-ent catalogues. Even if a large scatter is present, the mass discrepancy anti–correlates with the relative separation of the galaxies. Moreover, this trend does not depend on the adopted galaxy identification procedure. According to this evidence we interpret the discrepancy between LAP method and $n$–body simulation of an $\Omega_o = 1$ CDM Universe as mainly due to the existence of extended overlapping halos centered on the visible galaxies. We have verified that the quasi-homogeneous intragroup background cannot be accurately modeled by varying the softening parameter. Since the LAP technique measures the clustered mass inside the orbits, then in an $\Omega_o = 1$ Universe this method can be safely applied to only a particular class of systems:

- The self gravitating system studied has to be dynamically young (P1).
- The average relative separations between objects has to be larger than $\sim 2$ Mpc, that is, much larger than the collapsed galaxies.

We conclude that the LAP method applied to Local Group cannot rule out an $\Omega_o = 1$ value on LG scales. However the LAP method remains a powerful technique for systems large compared to the collapse radius of galaxies, for instance, the infall to galaxy clusters.

We thank H.M.P. Couchman for providing $n$–body data. E.B. is grateful to P. Catelan, T. Kroeker, L. Moscardini and R. Van de Weygaert for fruitful discussions, S. Huang, O. Lopez-Cruz, I. Prandoni and especially J. Wadsley for their valuable help. E.B. also thanks SISSA for financial support and the Department of Astronomy of the University of Toronto for its hospitality during this work.

# REFERENCES

Barnes, J, & Hut, P. 1986, Nature, 324, 446

Bean, A. J., Efstathiou, G., Ellis, R. S., Peterson, B. A., & Shanks, T. 1983, *MNRAS* , 205, 605

Carlberg, R. G., & Couchman, H. M. P. 1989, *ApJ* , 340, 47

Carlberg, R. G., Couchman, H. M. P., & Thomas, P. A. 1990, *ApJ* , 352, L29

Carlberg, R. G. 1991, *ApJ* , 367, 385

Carlberg, R. G., & Dubinski, J. 1991, *ApJ* , 369, 13

Carlberg, R. G. 1993, *ApJ* , preprint

Cen, R., & Ostriker, J. P. 1992, *ApJ* , 399, L113

Couchman H. M. P., & Carlberg, R. G. 1992, *ApJ* , 389, 453

Davis, M., & Peebles. P. J. E. 1983, *ApJ* , 267, 465

Dekel, A., Bertschinger, E., Yahil, A. Strauss, M. A., Davis, M. & Huchra, J. P. 1993, *ApJ* , 412, 1

Dubinski, J. 1988, M. Sc. thesis. University of Toronto.dubinski

Dunn, A. M., & Laflamme. R. 1993, *MNRAS* , 426, 865

Einasto, J., & Lynden-Bell, D. 1982, in Proc. Study Week on Cosmology and Fundamental Physics, Astrophysical Cosmology, Vol. 48, ed. H. A. Bruck, G. V. Coyne & M. S. Longair (Rome: Specola Vaticana), 85

Faber, S. M., & Gallagher, J. S. 1979, *Ann. Rev. Astr. Ap.* , 17, 35

Gelb, J. M., & Bertschinger, E. 1993, preprint

Gott, J. R. III, & Thuan, T. X 1978, *ApJ* , 223, 426

Gunn. J. E. 1975, Comm. Ap.Space, Phys., 6, 7

Hale-Sutton, D., Fang, R., Metcalfe, N., & Shanks, T. 1989, *MNRAS*, 237, 569

Kahn, F. D., & Woltjer. L. 1959, *ApJ*, 130, 705

Kaiser, N., *et al.* 1991, *MNRAS*, 252, 1

Katz, N., & White, S. D. M. 1993, preprint

Kroeker, T. L., & Carlberg, R. G. 1991, *ApJ*, 376, 1

Little, B., & Tremaine. S. 1987, *ApJ*, 320, 493

Mc Call, M. L. 1989, *AJ*, 97, 1341

Mishra, R. 1985, *MNRAS*, 212, 163

Peebles, P. J. E. 1976, *ApJ*, 205, L109

Peebles, P. J. E. 1989, *ApJ*, 332, 17

Peebles, P. J. E. 1989, *ApJ*, 344, L53 (P1)

Peebles, P. J. E. 1990, *ApJ*, 362, 1 (P2)

Peebles, P. J. E. 1993, Principles of Physical Cosmology (Princeton: Princeton University Press)

Peebles, P. J. E. 1994, preprint

Raychaudhury, S., & Lynden-Bell. D. 1989, *MNRAS*, 240, 195

Rowan-Robinson, M. *et al.* . 1990, *MNRAS*, 247, 1

Salucci, P., Persic, M. & Borgani, S. 1993, *ApJ*, 405, 459

Strauss, M. A., & Davis, M. 1988, in Large-Scale Motions in the Universe, ed. V. C. Rubin & G. V. Coyne, S.J. (Princeton: Princeton Univ. Press), 255

Strauss, M. A., Yahil, A., Davis, M., Huchra, J. P., & Fisher, K. B.

1992, *ApJ*, 397, 395

Zatrisky, D., Olszewski, W., Shommer, A. R., Peterson, R. C., & Aaronson, M. 1989, *ApJ*, 345, 759

# FIGURE CAPTIONS

**Fig. 1.** Time evolution of a density peak identified at $t = 17.8$. The X-Y projections at times $t =$ 6.0, 10.5, and 17.8 are displayed. The merging processes clearly characterize its dynamical evolution.

**Fig. 2a/b.** X-Y projections of the orbits for galaxies selected at $\frac{\delta \rho}{\rho} = 125$ and $\frac{\delta \rho}{\rho} = 8000$ respectively. Dashed lines represent the true orbits from the $n$-body experiment. Continuous lines are the orbit found by the LAP method. Black dots are the final positions of the galaxies. The orbits are traced back to a redshift 7. The circles have a radius of 5 Mpc. LAP orbits are quite similar to the true orbits. Significant differences in the orbits of bodies 6 and 7 (fig 2a) and 12 (figure 2b) probably indicate that true orbits represent stationary points which are no minima for the action.

**Fig. 3a/b.** Peculiar velocity found using LAP method vs. true peculiar velocity. The numbers refer to the galaxies in figure 2a/b. The LAP method causes external galaxies to be less bound.

**Fig. 4a/b.** $M(< R)$ averaged over all the binaries vs. $R - r_{min}$ (in Mpc units). Binary galaxies selected at an overdensity threshold $\frac{\delta \rho}{\rho} = 8000$ using two different selection criteria are shown. The existence of extended haloes surrounding the binaries is inferred.

**Fig. 5a/b.** $\log(M_5/M_{LAP})$ vs. $r_1$. Binaries selected using two different overdensity threshold are shown. Black dots refers to 'wide' binary systems, filled triangles are for the binary included in the second category

and small dots refers to bodies belonging to the third category. The best LG candidates are contained in the the region delimitated by two vertical lines. In both case the mass underestimate anti–correlates with the relative separation of the galaxies

Table 1
n-body test parameters

| # (1) | $\delta\rho/\rho$ (2) | N (3) | $M_T/M_{LAP}$ (4) |
|---|---|---|---|
| 1 | 45 | 9 | 1.47 |
| 2 | 125 | 13 | 1.25 |
| 3 | 1000 | 13 | 2.11 |
| 4 | 2500 | 9 | 1.60 |
| 5 | 6000 | 14 | 1.60 |
| 6 | 43000 | 9 | 1.77 |
| 7 | 125000 | 10 | 1.30 |
| 8[a] | 8000 | 10 | 1.76 |

[a] 'two-steps' identification procedure.

Table 2
*n*-body test parameters

| # (1) | $\delta\rho/\rho$ (2) | $N_1$ (3) | Mean$_1$ (4) | $\sigma_1$ (5) | $N_2$ (6) | Mean$_2$ (7) | $\sigma_2$ (8) | $N_{LG}$ (9) | Mean$_3$ (10) | Mean$_4$ (11) |
|---|---|---|---|---|---|---|---|---|---|---|
| $1^{t1}$ |  | 32 | 0.5 | 0.2 | 24 | 0.5 | 0.2 |  |  |  |
| $2^{t2}$ | $125^a$ | 47 | 1.0 | 0.5 | 46 | 1.0 | 0.5 | 3 | 0.6 | 0.7 |
| $3^{t3}$ |  | 49 | 1.6 | 1.0 | 49 | 1.7 | 1.0 |  |  |  |
| $4^{t1}$ |  | 19 | 0.4 | 0.2 | 12 | 0.4 | 0.2 |  |  |  |
| $5^{t2}$ | $8000^a$ | 31 | 0.8 | 0.4 | 29 | 0.8 | 0.4 | 4 | 0.7 | 0.9 |
| $6^{t3}$ |  | 38 | 1.3 | 0.7 | 39 | 1.4 | 0.7 |  |  |  |
| $7^{t1}$ |  | 2 | 0.4 | 0.4 | 1 | 0.8 | - |  |  |  |
| $8^{t2}$ | $64000^a$ | 11 | 0.8 | 0.3 | 8 | 0.9 | 0.3 | 1 | 1.4 | 1.44 |
| $9^{t3}$ |  | 18 | 1.3 | 0.7 | 20 | 1.3 | 0.6 |  |  |  |
| $10^{t1}$ |  | 17 | 0.4 | 0.2 | 12 | 0.5 | 0.2 |  |  |  |
| $11^{t2}$ | $125^b$ | 33 | 0.7 | 0.4 | 32 | 0.8 | 0.4 | 2 | 0.65 | 0.87 |
| $12^{t3}$ |  | 39 | 1.1 | 0.7 | 40 | 1.3 | 0.7 |  |  |  |
| $13^{t1}$ |  | 5 | 0.3 | 0.2 | 2 | 0.5 | 0.2 |  |  |  |
| $14^{t2}$ | $8000^b$ | 13 | 0.6 | 0.3 | 11 | 0.8 | 0.4 | 0 | - | - |
| $15^{t3}$ |  | 17 | 1.1 | 0.7 | 18 | 1.2 | 0.7 |  |  |  |
| $16^{t1}$ |  | 1 | 0.6 | - | 1 | 0.7 | - |  |  |  |
| $17^{t2}$ | $64000^b$ | 4 | 0.6 | 0.2 | 4 | 0.6 | 0.3 | 0 | - | - |
| $18^{t3}$ |  | 11 | 0.7 | 0.3 | 18 | 0.7 | 0.3 |  |  |  |

$^a$'Single Step' Procedure.
$^b$'Double Step' Procedure.
$^{t1}$ Only 'wide' binaries are considered.
$^{t2}$ 'Wide' and 'quasi-wide' binaries are considered.
$^{t3}$ All binaries are considered.

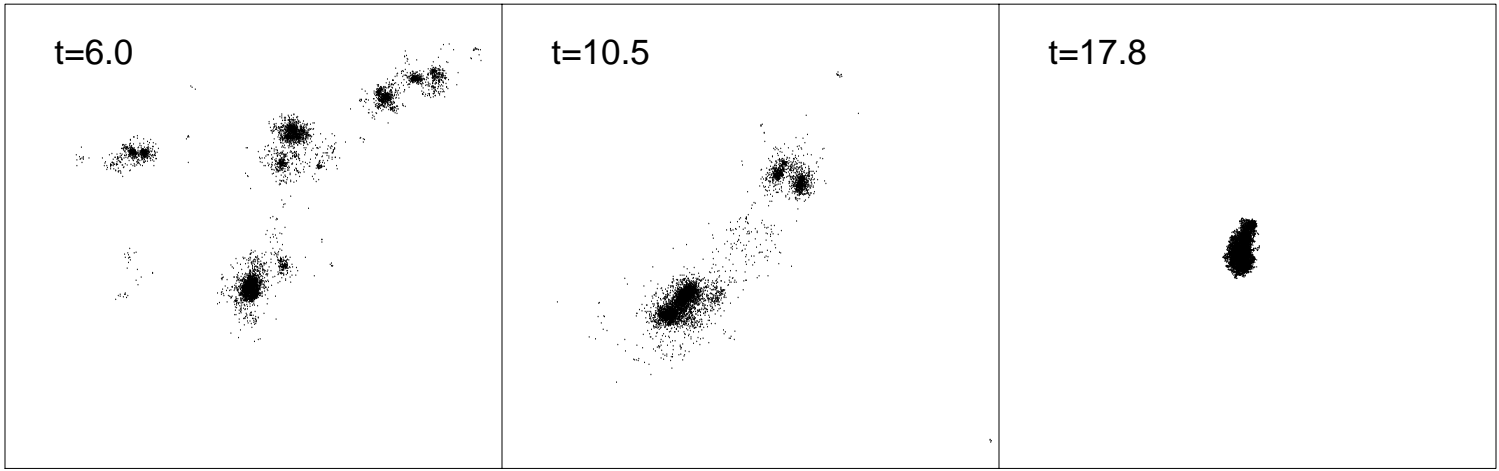

$\delta\rho/\rho = 8000$

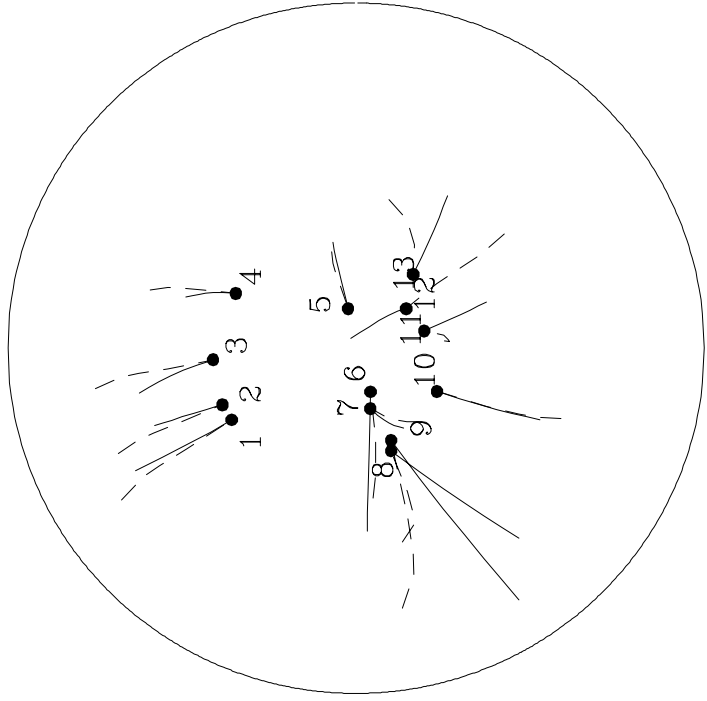
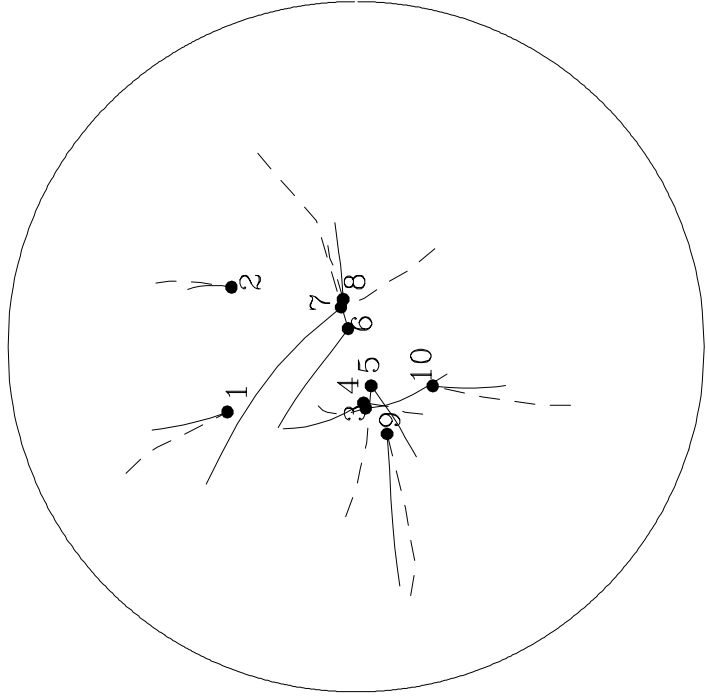

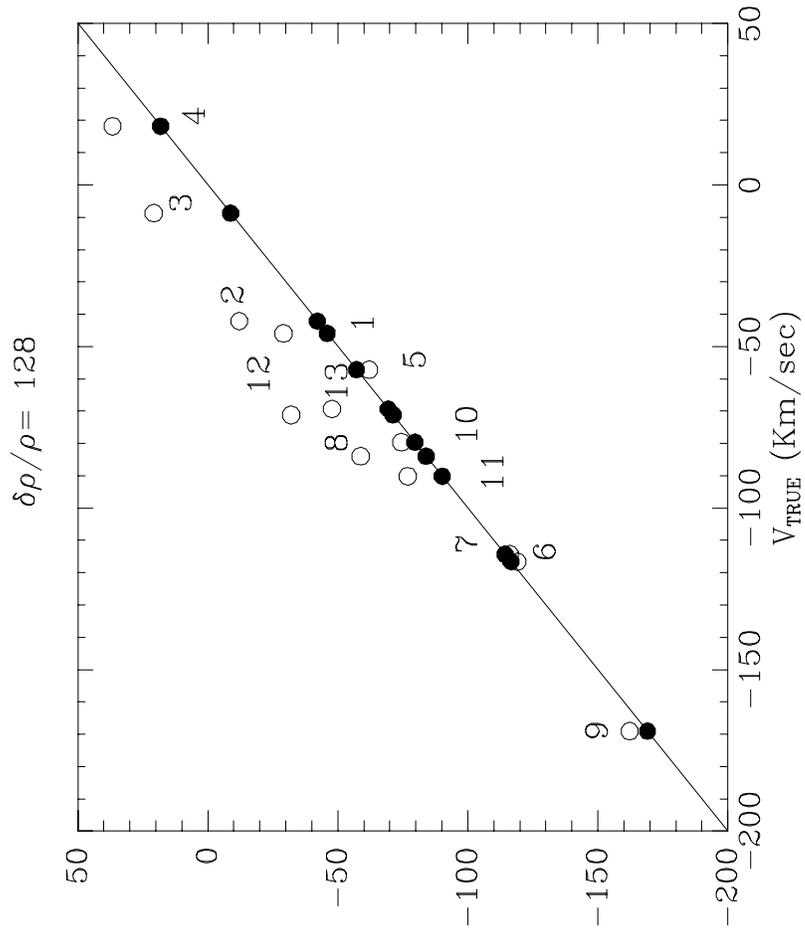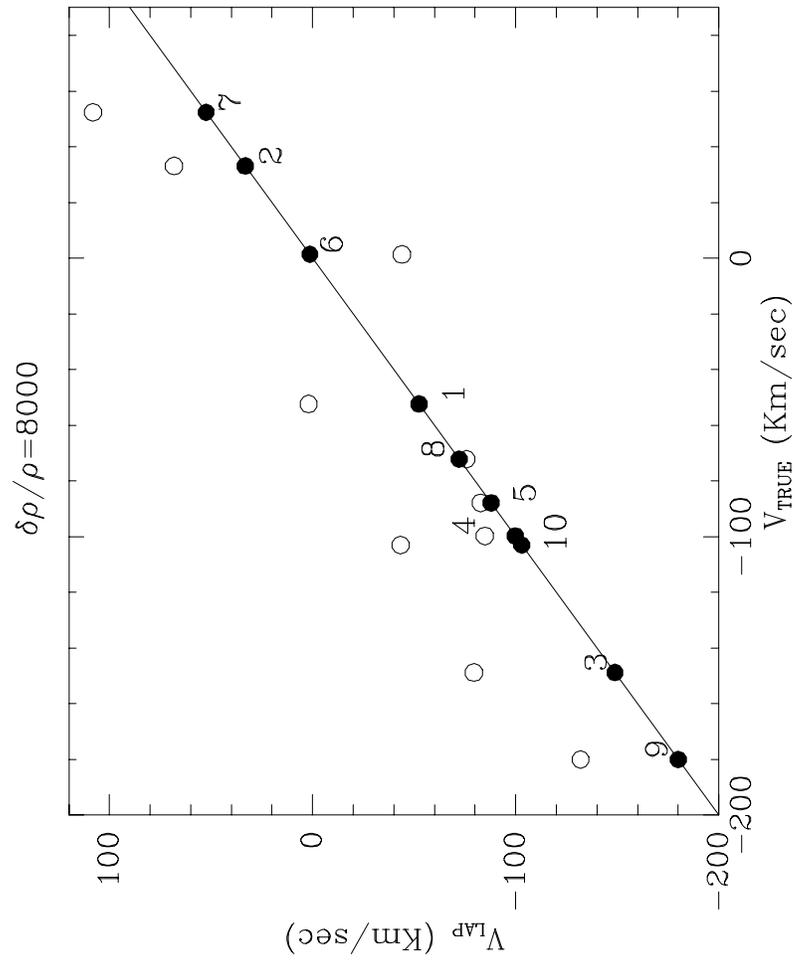

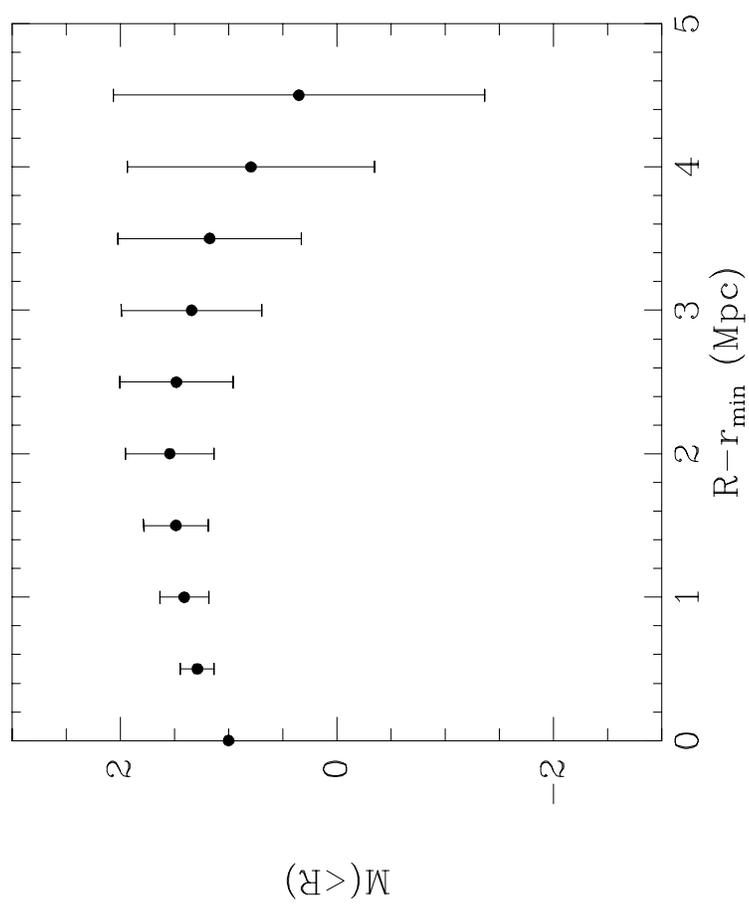

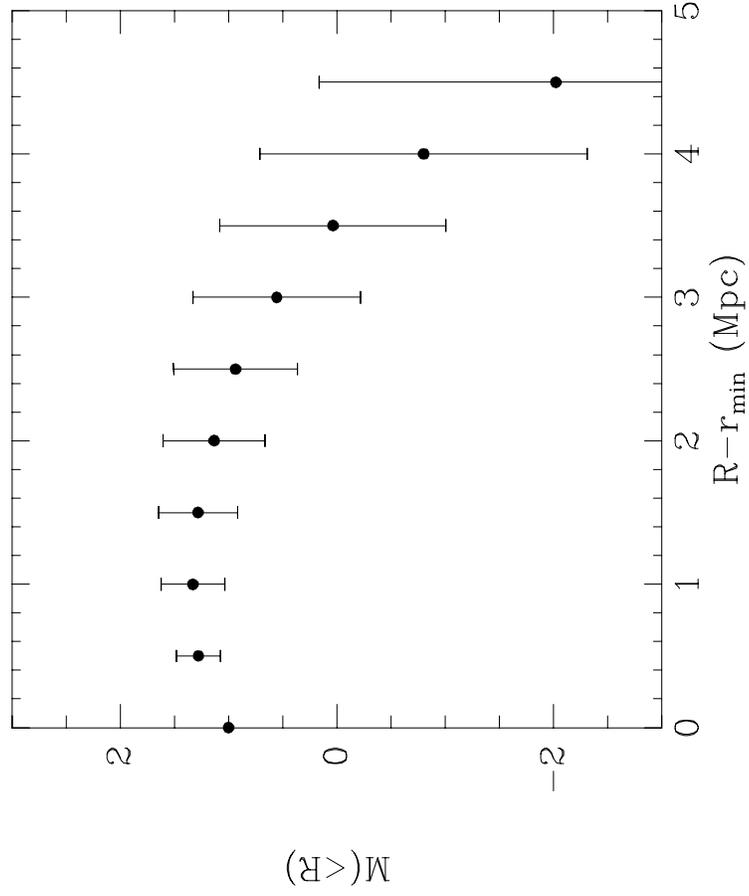

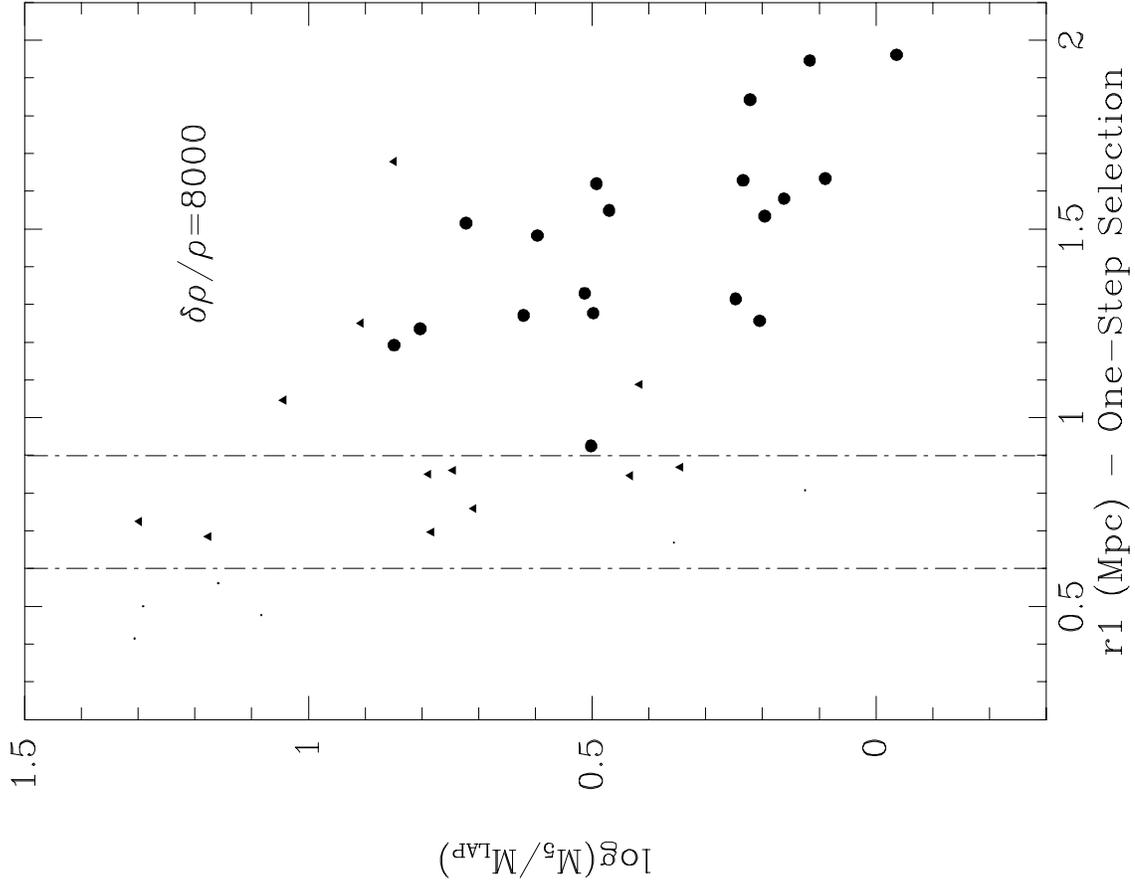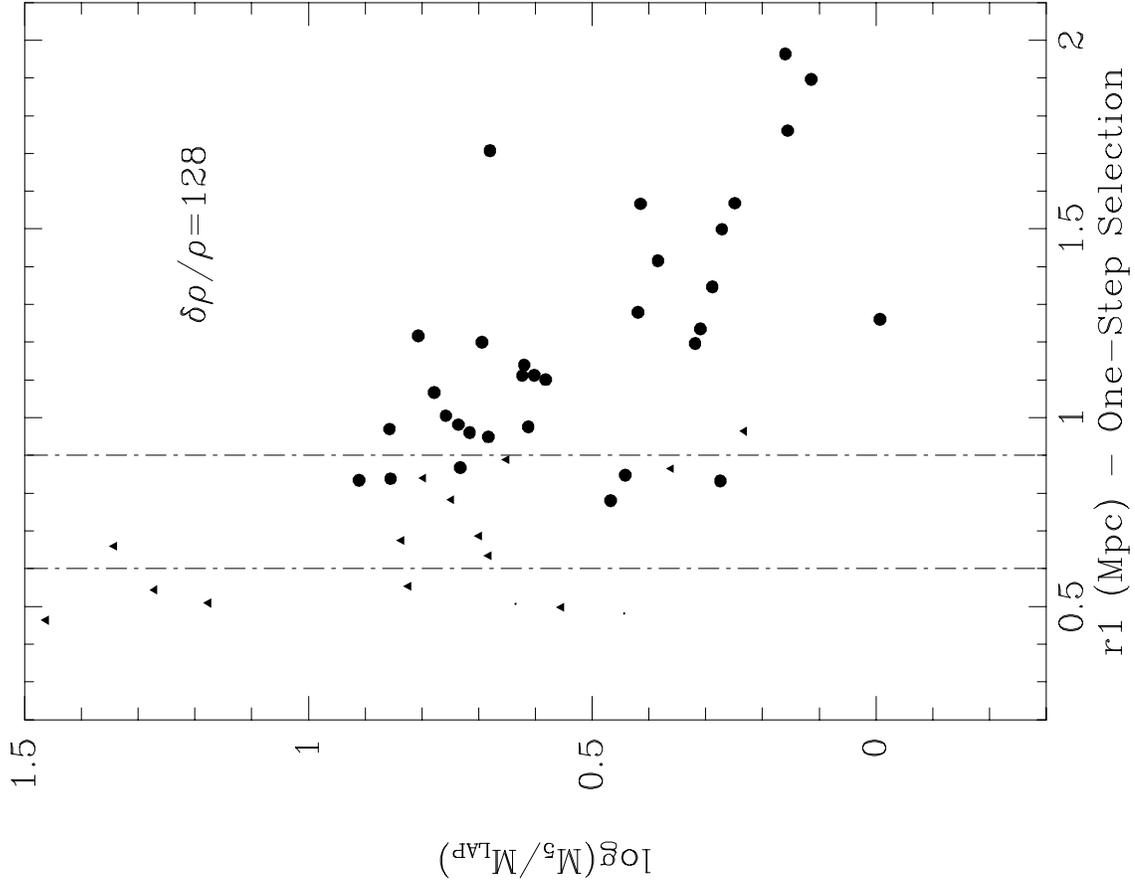